\documentclass[a4paper]{article}
\usepackage[utf8]{inputenc}
\usepackage[margin=1in]{geometry}     
\usepackage[parfill]{parskip}  
\usepackage{graphicx}
\usepackage{amssymb}
\usepackage{braket}
\usepackage{epstopdf}
\usepackage{mathtools}
\usepackage{subcaption}
\usepackage[toc,page]{appendix}
\usepackage{cases}
\usepackage{tikz} 
\usetikzlibrary{arrows} 
\usetikzlibrary{shapes,arrows,positioning,automata,backgrounds,calc,er,patterns}
\tikzset{
    ->-/.style={decoration={
  markings,
  mark=at position .5 with {\arrow{>}}},postaction={decorate}},
    -<-/.style={decoration={
  markings,
  mark=at position .5 with {\arrow{<}}},postaction={decorate}},
    ->/.style={decoration={
  markings,
  mark=at position .4 with {\arrow{>}}},postaction={decorate}},
}
\usetikzlibrary{shapes,arrows,positioning,automata,backgrounds,calc,er,patterns}
\usetikzlibrary{calc,positioning,shadows.blur,decorations.pathreplacing}
\tikzset{%
        brace/.style = { decorate, decoration={brace, amplitude=5pt} },
       mbrace/.style = { decorate, decoration={brace, amplitude=5pt, mirror} },
        label/.style = { black, midway, scale=0.7, align=center },
     toplabel/.style = { label, above=.5em, anchor=south },
    leftlabel/.style = { label,rotate=-90,left=.5em,anchor=north },   
  bottomlabel/.style = { label, below=.5em, anchor=north },
        force/.style = { rotate=-90,scale=0.6 },
        round/.style = { rounded corners=2mm },
       legend/.style = { right,scale=0.6 },
        nosep/.style = { inner sep=0pt },
   generation/.style = { anchor=base }
}

\usepackage{tikz-feynman}
\usepackage{empheq}
\usepackage{slashed}
\usepackage{changepage}
\usepackage{jheppub}
\usepackage{fontawesome}

\newcommand{\be}{\begin{equation}}
\newcommand{\ee}{\end{equation}}
\newcommand{\ba}{\begin{eqnarray}}
\newcommand{\ea}{\end{eqnarray}}

\preprint{ \vbox{\hbox{IPARCOS-UCM-26-033}}}

\title{The left-cut for partial waves in terms of physical amplitudes}
\author{Alexandre Salas-Bernárdez}
\affiliation{Dept.  Análisis Matemático y Matemática Aplicada, Univ. Complutense de Madrid, Plaza de las Ciencias 3, 28040 Madrid, Spain. (On leave at University of California San Diego).}
\emailAdd{alexsala@ucm.es}

\abstract{We derive a novel representation of the partial wave amplitude over the left-hand cut for $2 \to 2$ scattering. We express the left-hand cut of arbitrary isospin and angular momentum partial waves as an integral of right-hand cut imaginary parts. This formulation provides an explicit, exact extraction of the logarithmic branch cut structures, offering a valuable tool to systematically quantify left-hand cut uncertainties in unitarization methods such as the Inverse Amplitude Method or $N/D$ approaches.}

\begin{document}

\maketitle

\section{Introduction}\label{sec1}
The Effective Field Theory (EFT) approach has become standard in many applications in particle physics, with prominent successes in hadron physics and in the quest for unveiling New Physics at high energies. Both the physics of light hadrons and the dynamics of the Electroweak Symmetry Breaking Sector (EWSBS) can be systematically described using Chiral Perturbation Theory (ChPT) \cite{Weinberg:1978kz, Gasser:1983yg}. In the EWSBS, additional dynamics modifying the known interactions of the Standard Model (SM) are encoded in higher-dimensional effective operators within the Higgs EFT (HEFT) \cite{Brivio:2016fzo} or the Standard Model EFT (SMEFT) \cite{Jenkins:2013zja,Brivio:2017vri,Alioli:2022fng}. With the former being more general than the latter, as shown  theoretically and phenomenologically in \cite{Alonso:2016oah,Cohen:2020xca,Gomez-Ambrosio:2022qsi} and \cite{Delgado:2023ynh,Domenech:2025gmn}.

However, EFTs are intrinsically limited by their perturbative unitarity breakdown at high energies. To extend their predictive power to the resonance region, one must rely on dispersion relations and unitarization techniques. It is well known from the application of exact $S$-matrix methods---such as Roy equations \cite{Roy:1971tc}, Garcia-Martin--Kaminski--Pelaez--Yndurain (GKPY) equations \cite{Garcia-Martin:2011iqs}, and the Froissart-Gribov representation \cite{Froissart:1961ux, Gribov:1961ex}---that the unphysical left-hand cut is entirely dictated by crossing symmetry from the physical channels. 

In this work, we develop an explicit algebraic series representation that connects, through crossing, the left-hand cut of a partial wave directly with the amplitude evaluated at its physical $t$- and $u$-channels. This representation appears to be robust numerically for the NLO $SU(2)$ isospin amplitudes, whereas the known Madelstam-Chew representation of Eq (IV.7) in \cite{Chew:1960iv} shows numerical instability as studied in \cite{Salas-Bernardez:2020hua}.

This representation may be particularly useful for constraining the theoretical uncertainties of unitarization methods such as the Inverse Amplitude Method (IAM) \cite{Dobado:1996ps}, the improved $K$-matrix method \cite{Truong:1991gv,Chung:1995dx}, or the $N/D$ method \cite{Oller:1998zr}. Since all these methods make distinct approximations regarding the analytic structure of the left-hand cut, a rigorous analytic control of their uncertainty is crucial for the robust prediction of new physics scales \cite{Salas-Bernardez:2020hua} produced by resonances in the EWSBS \cite{Delgado:2013hxa,Delgado:2015kxa}. In fact, the uncertainty in predicting a resonance's mass from low-energy data using the IAM is strongly dominated by its approximation of the left-cut for a given partial wave \cite{Salas-Bernardez:2020hua}. Once the left-cut expression is obtained, this uncertainty can be computed using Eq. (14) of \cite{Salas-Bernardez:2020hua}.

This short note is organized as follows, in Section \ref{mainsection} we obtain the representation of partial waves over the left cut for arbitrary angular momentum and isospin,
and, in Section \ref{sec:iso2}, we explicitly obtain some partial waves expressions over the left cut for $SU(2)$ isospin symmetric theories.

\section{Partial-wave amplitude over the left cut}\label{mainsection}
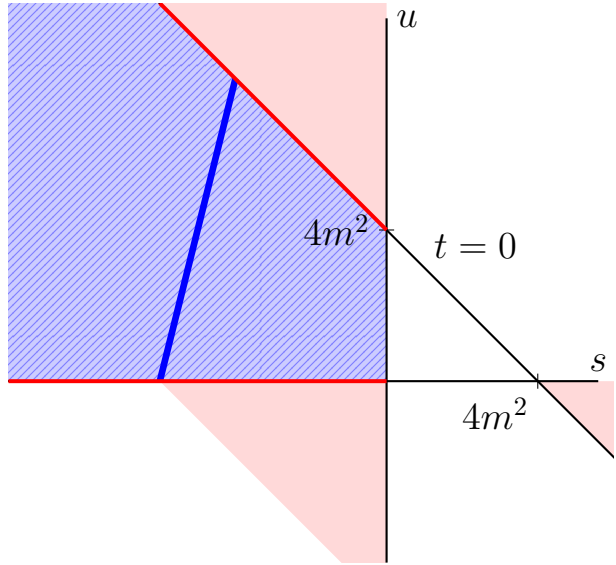
\begin{figure}[ht!]
\centering
\begin{tikzpicture}[scale=2]

    \clip (-2.5, -1.2) rectangle (1.5, 2.5);

    \fill[red!15] (1, 0) -- (3, -2) -- (3, 0) -- cycle;

    \fill[red!15] (0, 1) -- (-3, 4) -- (0, 4) -- cycle; 

    \fill[red!15] (0, 0) -- (0, -1.5) -- (-1.5, 0) -- cycle;

    \fill[blue!20] (-3, 0) -- (0, 0) -- (0, 1) -- (-3, 4) -- cycle;

    \fill[pattern=north east lines, pattern color=blue!50] (-3, 0) -- (0, 0) -- (0, 1) -- (-3, 4) -- cycle;

    \draw[red!80!black, thick] (-3, 0) -- (0, 0);       %
    \draw[red!80!black, thick] (-3, 4) -- (0, 1);       %
    \draw[thick] (-2.5, 0) -- (1.4, 0) node[above, font=\Large] {$s$};
    \draw[thick] (0, -1.2) -- (0, 2.4) node[right, font=\Large] {$u$};

    \draw[thick] (-2.5, 3.5) -- (2.5, -1.5) node[pos=0.55, above right, font=\Large] {$t = 0$};

    \draw (1, 0.05) -- (1, -0.05) node[below left, font=\Large] {$4m^2$};
    \draw (0.05, 1) -- (-0.05, 1) node[left, font=\Large] {$4m^2$};

    \draw[blue, line width=2.5pt] (-1.5, 0) -- (-1, 2);

    \draw[red, line width=1.5pt] (0, 1) -- (-1.5, 2.5);

\draw[red, line width=1.5pt] (0, 0) -- (-2.5, 0);

\end{tikzpicture}
\caption[Mandelstam plane for two identical particles]{
Mandelstam plane for $2 \to 2$ scattering of identical particles ( with $t = 4m^2 - s - u$). The three physical regions (misty rose) extend outward from the triangle vertices. The unphysical integration domain for the left cut (patterned purplish) is traversed by the arc (thick blue line) corresponding to $x \in [-1, 1]$. Endpoints lie on the $t$- and $u$-channel physical cuts (red).
}
\label{fig:mandelstam}
\end{figure}

In this section we will be dealing with the $2\to2$  Goldstone boson scattering amplitude of isospin $I$ in terms of the Mandelstam variables $s$, $t$ and $u$. In accordance with Mandelstam's hypothesis, we postulate the existence of a single, unique analytic amplitude $T(s,t,u)$ that evaluates to the respective channel amplitudes in their physical kinematic regimes:$$T(s,t,u) = \left\{
\begin{array}{l}
T_s(s,t,u) \quad \text{for} \quad s\geq 4m^2,\,t\leq 0,\,u\leq 0\ , \\
T_t(t,s,u) \quad \text{for} \quad t\geq 4m^2,\,s\leq 0,\,u\leq 0\ , \\
T_u(u,t,s) \quad \text{for} \quad u\geq 4m^2,\,t\leq 0,\,s\leq 0\ .
\end{array}
\right.$$ These three distinct physical regions corresponding to the different channels are illustrated by the shaded rose-colored areas in Fig.~\ref{fig:mandelstam}.

The partial wave projection along the left-hand cut ($s \leq 0$) is given by:

\begin{equation}
t_{J}^I(s) = \frac{1}{32\pi\eta} \int_{-1}^{+1} dx \, P_J(x) \, T^I(s,t(s,x),u(s,x)) \label{lcpartialwv}
\end{equation}
with the Mandelstam variables parametrized as $t(s,x) = (2m^2 - s/2)(1 - x)$ and $u(s,x) = (2m^2 - s/2)(1 + x)$ (here $x=\cos \theta$ with $\theta\in[0,\pi]$ being the scattering angle), and $P_J(x)$ being the Legendre polynomials of angular momentum $J$ ($\eta$ is the symmetry factor). As explained below, the integration domain in eq.~\eqref{lcpartialwv} corresponds to a line that traverses the unphysical region of the Mandelstam plane (purplish), ending on the physical $t$- and $u$-channel regions in Fig. \ref{fig:mandelstam}.

Let us define the reduced amplitude, $A(s,t(s,x))$ by expressing $u$ as a function of $s$ and $t$ (which is a function of $s$ and $x$ itself) by $u=4m^2-s-t$ (omitting isospin indices):
\begin{equation}
    A(s,t(s,x))=T(s,t(s,x),4m^2-s-t(s,x)).
\end{equation}

At the endpoints of the integral in eq. (\ref{lcpartialwv}), the amplitude evaluates to values on the physical right-hand cuts of the $t$- and $u$-channels:
\[
\begin{cases}
A(s,t(s,x=+1)) = T(s, 0, 4m^2 - s) = T_u(4m^2 - s, 0, s) \,, \\
A(s,t(s,x=-1)) = T(s, 4m^2 - s, 0) = T_t(4m^2 - s, s, 0) \,.
\end{cases}
\]
Since we evaluate $T(s+i\epsilon,x)$, the corresponding $t$ and $u$ arguments lie just below their respective right-hand cuts (see Eq. (\ref{ieprescript}) below).

The partial wave can be expressed in terms of auxiliary moments:
\begin{equation}
\psi_J(s) \equiv \int_{-1}^{+1} dx \, x^J \, T(s,t(s,x),u(s,x)) \,. \label{si}
\end{equation}
We partition the integration domain of eq. (\ref{si})  into two distinct segments: $x \in [-1, 0]$ and $x \in [0, +1]$. Within each subdomain, we expand the reduced amplitude $A(s,t(s,x))$ around the respective physical endpoints, $x = -1$ and $x = +1$. This procedure fundamentally assumes that both the real and imaginary parts of the scattering amplitude remain analytic in a neighborhood extending just below the physical $t$- and $u$-channel branch cuts. This assumption is justified by the fact that the scattering amplitude can be analytically continued across these physical cuts into the unphysical Riemann sheets. Furthermore, an inspection of the explicit ChPT expressions in App. \ref{app} confirms this property; the algebraic functions defining the imaginary parts along the cuts are themselves locally analytic functions of the Mandelstam variables and the scattering angle $x$ away from the threshold branch points.

In this way, integrating  eq. (\ref{si}), we find
\begin{align}
\psi_J(s) &= \sum_{n=0}^\infty \frac{(-1)^n J!}{(n+1+J)!} \Bigg[
\left. \frac{d^n A(s,t(s,x))}{d x^n} \right|_{x = +1} 
+ (-1)^{n+J} \left. \frac{d^n A(s,t(s,x))}{d x^n} \right|_{x = -1} 
\Bigg] \,.
\label{series}
\end{align}
This relation holds provided $s \neq 0$, to avoid the branch points from the $u$- and $t$-channels at $x = \pm1$. 

Let us notice now that 

\begin{equation}
   \left.  \frac{d A(s, t(s,x))}{d x}=(s/2-2m^2)\left(\frac{\partial T(s,t,u)}{\partial t} -\frac{\partial T(s,t,u)}{\partial u}\right)\right|_{\substack{t=t(s,x)\;\;\;\;\;\;\;\;\;\;\;\;\;\;\\
   u=4m^2-s-t(s,x)}}
\end{equation}
and that
\begin{equation}
    \left.\frac{d A(s,t)}{d t}=\left(\frac{\partial T(s,t,u)}{\partial t} +\frac{\partial u}{\partial t} \frac{\partial T(s,t,u)}{\partial u}\right)\right|_{u=4m^2-s-t}
\end{equation}
so that, since $\partial u/\partial t=-1$,
\begin{equation}
    \frac{d A(s, t(s,x))}{d x}=(s/2-2m^2)\frac{d A(s,t)}{d t}\;.
\end{equation}
Which generalizes to
\begin{equation}
    \frac{d^n A(s, t(s,x))}{d x^n}=(s/2-2m^2)^n\frac{d^n A(s,t)}{d t^n},
\end{equation}
with the $u$ derivative of $A$ being obtained in a similar fashion.

Hence, we can now identify the endpoint $x$-derivatives in eq. (\ref{series}) with those of the amplitudes in the $u$- and $t$-channels:
\begin{align}
\left. \frac{d^n A(s,t(s,x))}{d x^n} \right|_{x=+1} 
&= (2m^2 - s/2)^n \left. \frac{d^n T(s,t(s,u),u)}{d u^n} \right|_{u = 4m^2 - s} \,, \\
\left. \frac{d^n A(s,t(s,x))}{d x^n} \right|_{x=-1} 
&= (s/2 - 2m^2)^n \left. \frac{d^n T(s,t,u(s,t))}{d t^n} \right|_{t = 4m^2 - s} \,.
\end{align}

Now let us use the relation derived in \cite{Manohar:2008tc} that computes the derivatives of the scattering amplitude over the right cut (here $s$ stands for $s+i\epsilon$ and repeated indices are summed over):

\begin{equation}
\frac{d^n}{d s^n} T^{I}(s,t,u(s,t))
=
\frac{n!}{\pi}
\int_{4m^2}^{\infty} ds'
\left[
\frac{\delta^{II'}}{(s'-s)^{n+1}}
+
(-1)^n
\frac{C^{II'}_{u}}{(s'-u)^{n+1}}
\right]
\operatorname{Im} T^{I'}(s'+i\epsilon,t,u).\label{relationManohar}
\end{equation}
valid for $s>4m^2$ and $-28m^2 \le t \le 4m^2$ at least. Here $u(s,t)=4m^2-s-t$.

A comment on the $i\epsilon$ prescription is due here to get the correct imaginary parts on crossed channels. The kinematical condition, $4m^2=s+t+u$,  imposes that the imaginary part of $s$ is compensated with the imaginary parts of $t$ and $u$. And, to use the crossed relation in eq. (\ref{relationManohar}) above, we need $t$ or $u$ above their respective right cuts. To do so we will use Schwarz reflection principle to relate these as
\begin{equation}
T^I(s+i\epsilon,t-i\epsilon', u(s+i\epsilon,t-i\epsilon '))= \left(C_{t}^{I I'}T^{I'}(t+i\epsilon',s-i\epsilon, u(s-i\epsilon,t+i\epsilon ') \right)^*,\label{ieprescript}
\end{equation}
with $\epsilon'=\epsilon (1-x)/2$. One can check that $u$ gets modified as $u-i\epsilon''$ with $\epsilon''=\epsilon (1+x)/2$.

In this way, we have the following relations below the $t$-channel cut (omitting the $i\epsilon$ for brevity):

\begin{align}
    &\left(\frac{d^n T^{I}(s,t,u(s,t))}{d t^n} \right)^*=C_t^{II'}\frac{d^n T^{I'}(t,s,u)}{d t^n} =\nonumber\;\;\;\\
    &\;\;\;=C_t^{II'}\left(\frac{n!}{\pi}
\int_{4m^2}^{\infty} ds'
\left[
\frac{\delta^{I'I''}}{(s'-t)^{n+1}}
+
(-1)^n
\frac{C^{I'I''}_{u}}{(s'-u)^{n+1}}
\right]
\operatorname{Im} T^{I''}(s'+i\epsilon,s,u)\right),
\end{align}
valid for $t\geq 4m^2$ and $-28m^2 \le s \le 4m^2$ at least. Here $u(s,t)=4m^2-s-t$.

For the amplitude and its derivatives below the $u$-channel cut we derive analogous relations:
\begin{align}
    &\;\;\left(\frac{d^n T^{I}(s,t(s,u),u)}{d u^n} \right)^*=C_t^{II'}C_u^{I'I''}\frac{d^n T^{I''}(u,s,t)}{d u^n} =\nonumber\\
    &\;\;\;\;=C_t^{II'}C_u^{I'I''}\left(\frac{n!}{\pi}
\int_{4m^2}^{\infty} ds'
\left[
\frac{\delta^{I''I'''}}{(s'-u)^{n+1}}
+
(-1)^n
\frac{C^{I''I'''}_{u}}{(s'-t)^{n+1}}
\right]
\operatorname{Im} T^{I'''}(s'+i\epsilon,s,t)\right),
\end{align}
valid for $u\geq 4m^2$ and $-28m^2 \le u \le 4m^2$ at least. Here $t(s,u)=4m^2-s-u$.

Then, using that $C_u^{II'}C_u^{I'I''}=\delta^{II''}$ and evaluating at $x=\pm 1$ we find:
\begin{align}
    &\left.\frac{d^n T^{I}(s,t,u)}{d u^n} \right|_{x = +1}=\frac{n!}{\pi}
\int_{4m^2}^{\infty} ds'
\left[
\frac{C_t^{II'}C_u^{I'I''}}{(s'-4m^2+s)^{n+1}}
+
(-1)^n
\frac{C^{II''}_{t}}{(s')^{n+1}}
\right]^*
\operatorname{Im} T^{I''}(s'+i\epsilon,s,0),\\
    &\left.\frac{d^n T^{I}(s,t,u)}{d t^n} \right|_{x = -1} =\frac{n!}{\pi}
\int_{4m^2}^{\infty} ds'
\left[
\frac{C_t^{II''}}{(s'-4m^2+s)^{n+1}}
+
(-1)^n
\frac{C_t^{II'}C^{I'I''}_{u}}{(s')^{n+1}}
\right]^*
\operatorname{Im} T^{I''}(s'+i\epsilon,s,0).
\end{align}
Using this information in eq. (\ref{series}) we obtain (assuming real crossing matrices)

\begin{align}
\psi_J(s) &= ( C_t^{II'}C_u^{I'I''}+ (-1)^J C_t^{II''})\sum_{n=0}^\infty \frac{(-1)^n J!}{(n+1+J)!} \frac{n!}{\pi}\times\nonumber\\
&\times(2m^2 - s/2)^n\int_{4m^2}^{\infty} ds' \left[
\frac{1}{(s'-4m^2+s)^{n+1}}
+
\frac{(-1)^{n+J}}{(s')^{n+1}}
\right]^* \operatorname{Im} T^{I''}(s'+i\epsilon,s,0).
\end{align}

We can sum this infinite series using Mathematica 14.2 \cite{Mathematica}. 

In this way, for $J=0$ one obtains
\begin{align}
t_0^I(s) &= \frac{2}{\pi}\frac{( C_t^{II'}C_u^{I'I''}+  C_t^{II''})}{{4 m^2-s} }\int_{4m^2}^{\infty} ds'\, {\left[ \log \left(\frac{s'}{ s'+s-4m^2}\right)\right]^*}\operatorname{Im} T^{I''}(s'+i\epsilon,s,0).
\end{align}

For $J=1$:
\begin{align}
t_1^I(s) &= -\frac{2}{\pi}\frac{ (C_t^{II'}C_u^{I'I''}-  C_t^{II''})}{{4 m^2-s}} \times\nonumber\\
&\times\int_{4m^2}^{\infty} ds' \left[2+ \log \left(\frac{s'}{ s'+s-4m^2}\right)\right.\left.-2 \frac{s'}{4m^2-s} \log \left(\frac{s'}{ s'+s-4m^2}\right)\right]^*\operatorname{Im} T^{I''}(s'+i\epsilon,s,0).
\end{align}

And for $J=2$:
\begin{align}
t_2^I(s) &= \frac{2}{\pi}\frac{ (C_t^{II'}C_u^{I'I''}+  C_t^{II''})}{{4 m^2-s}} \int_{4m^2}^{\infty} ds'\left[ 3+\log \left(\frac{s'}{s'+s-4m^2}\right)+6 \left(\frac{s'}{4m^2-s}\right)^2 \log \left(\frac{ s'}{s'+s-4m^2}\right)\right.\nonumber\\
& \left.-6 \frac{s'}{4m^2-s} \left(\log \left(\frac{s'}{s'+s-4m^2}\right)+1\right)\right]^*\operatorname{Im} T^{I''}(s'+i\epsilon,s,0).
\end{align}
Interestingly, for arbitrary $J$, these integrals will converge as long as the $\operatorname{Im} T^{I}(s'+i\epsilon,s,0)$ factors grow slower than $(s')^{J+1}$ for any $J$. We have explicitly checked this behavior up to $J=8$ and identified that the factors inside brackets behave as $\frac{(n!)^2(4m^2-s)^{J+1}}{(n+1)!(s')^{J+1}}$ when $s'\to +\infty$.

\subsection{Left-cut imaginary parts for $SU(2)$ isospin}\label{sec:iso2}

Now we are ready to compute the  imaginary part of the partial-wave amplitude over the left cut. For doing so we will use the distributional identity $\log{(x+i\epsilon)}=\log{|x|}+i\pi \Theta(-x)$ to find that, for $s'\to s'+i\epsilon$ and $s\to s+i\epsilon$ with $s<0$ and $s'\geq 4m^2$, we have
\begin{equation}
    \log{\frac{s'}{s'-(4m^2-s)}}=\log{\frac{s'}{|s'-(4m^2-s)|}}-i\pi \Theta\left(-(s'-(4m^2-s))\right)\;.
\end{equation}

\begin{figure}[ht!]
    \centering
    \includegraphics[width=\linewidth]{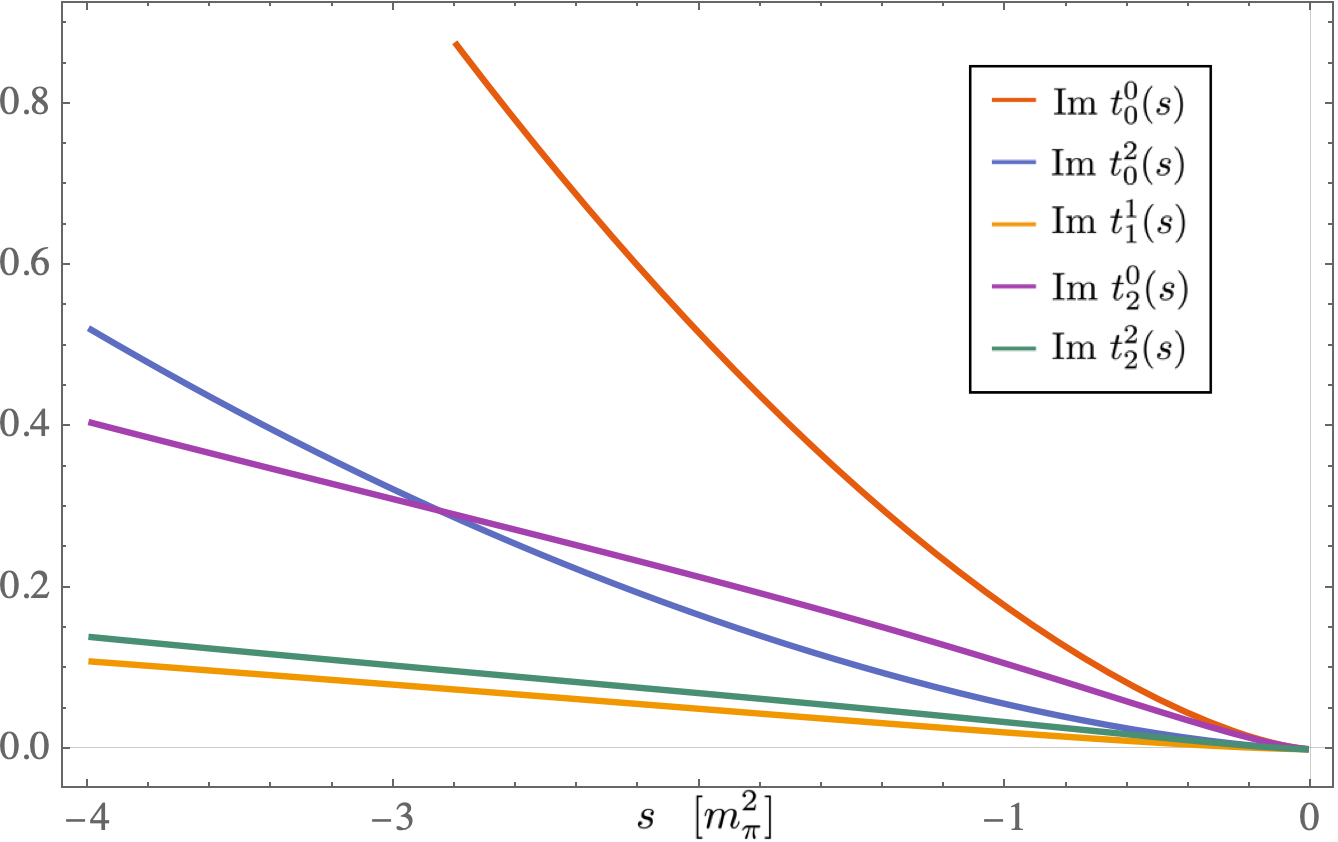}
    \caption{Imaginary parts of NLO $SU(2)$ isospin partial-wave amplitudes over the left cut. Here $s$ has units of $m_\pi^2$.}
    \label{fig:leftcuts}
\end{figure}

So that for $s<0$ over the left cut and for $SU(2)$ isospin Goldstone boson scattering (see App. \ref{app}) we have, for $I=0$ and $J=0$, 
\begin{align}
    \textrm{Im } t^0_0(s)=\frac{\frac{8}{3}\delta^{0I}+2\delta^{1I}+\frac{10}{3}\delta^{2I}}{{4 m^2-s} }\int_{4m^2}^{4m^2-s} ds' \operatorname{Im} T^{I}(s'+i\epsilon,s,0).
\end{align}
For $I=2$ and $J=0$,
\begin{align}
    \textrm{Im } t^2_0(s)=\frac{\frac{2}{3}\delta^{0I}-\delta^{1I}+\frac{7}{3}\delta^{2I}}{{4 m^2-s} }\int_{4m^2}^{4m^2-s} ds' \operatorname{Im} T^{I}(s'+i\epsilon,s,0).
\end{align}
For $I=1$ and $J=1$,
\begin{align}
    \textrm{Im } t^1_1(s)=-\frac{\frac{2}{3}\delta^{0I}+{3}\delta^{1I}-\frac{5}{3}\delta^{2I}}{{4 m^2-s} }\int_{4m^2}^{4m^2-s} ds' \left[ 1-\frac{2s'}{4m^2-s}\right]\operatorname{Im} T^{I}(s'+i\epsilon,s,0).
\end{align}
For $I=0$ and $J=2$
\begin{align}
    \textrm{Im } t^0_2(s)=\frac{\frac{8}{3}\delta^{0I}+2\delta^{1I}+\frac{10}{3}\delta^{2I}}{{4 m^2-s} }\int_{4m^2}^{4m^2-s} ds' \left[ 1-\frac{6s'}{4m^2-s}+6\left(\frac{s'}{4m^2-s}\right)^2\right]\operatorname{Im} T^{I}(s'+i\epsilon,s,0),
\end{align}
and, finally, for $I=2$ and $J=2$
\begin{align}
    \textrm{Im } t^2_2(s)=\frac{\frac{2}{3}\delta^{0I}-\delta^{1I}+\frac{7}{3}\delta^{2I}}{{4 m^2-s} }\int_{4m^2}^{4m^2-s} ds' \left[ 1-\frac{6s'}{4m^2-s}+6\left(\frac{s'}{4m^2-s}\right)^2\right]\operatorname{Im} T^{I}(s'+i\epsilon,s,0).
\end{align}
In Fig. \ref{fig:leftcuts} we plot the results for the left cuts of the imaginary parts for all partial wave amplitudes listed above.

It is interesting to notice how the $t_{1}^{1}(s)$ partial wave has the smallest imaginary part on the left cut compared to other partial waves. This may be a reason why the IAM works very well for the $IJ=11$ channel and fails for the $IJ=00$ channel (which has the largest left cut imaginary part) in predicting the position of a resonance, since the IAM amplitude is approximating the left-cut imaginary part imposing the perturbative unitarity condition of the right cut \cite{Salas-Bernardez:2020hua}.

\newpage

\section{Conclusions}

In this short note, we have derived a novel algebraic representation of the left-hand cut for partial-wave amplitudes. By relying on crossing symmetry, and a fixed-$t$ dispersion relation \cite{Manohar:2008tc}, we have expressed the left-hand cut entirely in terms of amplitudes in physical regions.  This relation holds for negative values of $s$ in the range $-28m^2 \le s < 0$ at least. In future work, we will compare this representation with the Mandelstam-Chew representation \cite{Chew:1960iv} of partial waves over the left cut. Nonetheless, practical applications of the Mandelstam-Chew representation have failed \cite{Salas-Bernardez:2020hua}, whereas the representation here obtained offers numerically stable solutions.

We have explicitly computed the imaginary parts of the partial waves over the left cut for the $IJ = 00, 02, 11, 20,$ and $22$ channels in an $SU(2)$ isospin-symmetric theory. Crucially, our exact analytic extraction reveals a quantitative  difference in the imaginary parts across different channels: the vector channel ($IJ=11$) exhibits the smallest imaginary part on the left cut, whereas the scalar and tensor channels present greater imaginary parts over the left cut. 

This observation may provide theoretical justification for the phenomenological behavior of unitarization methods. For instance, the standard Inverse Amplitude Method approximates the left-hand cut by utilizing the perturbative unitarity condition and the Next-to-Leading Order  expressions evaluated over the right cut. The  left-cut behavior of partial waves derived here may explain why the IAM succeeds remarkably well in predicting resonances in the $IJ=11$ channel (such as the $\rho$ meson) but faces systematic challenges in the $IJ=00$ channel (such as the $f_0(500)$ or $\sigma$ meson). Ultimately, this formalism offers a robust mathematical tool to systematically quantify and control left-hand cut uncertainties in effective field theories and unitarization techniques. This is so because, once a unitarization technique is used for the right cuts of crossed channels, the systematic uncertainty of the method on the right cut can straightforwardly be translated to the uncertainty in approximating the left-cut. We leave this endeavor for a future work.

\section*{Acknowledgments}
I would like to thank Professor Aneesh Manohar for very fruitful discussions and for hosting me at UCSD, as well as the UCSD Physics research staff for creating a welcoming atmosphere. I also acknowledge Professor Juan Ferrera and Professor Felipe Llanes-Estrada for their valuable discussions. Funded by research grant PID2022-137003NB-I00 from Spanish MCIN/AEI/10.13039/501100011033/  and EU FEDER.

\section*{Declaration of generative AI and AI-assisted technologies in the manuscript preparation process}

During the preparation of this work, the author used Google Gemini 3.1 Pro \cite{gemini31pro2026} for grammar/syntax checks and to obtain the expressions of the $SU(2)$ amplitude imaginary parts and crossing matrices from \cite{Gasser:1983yg}. The author reviewed and edited the output as needed and take full responsibility for the content of the published article.

\appendix
\newpage
\section{SU(2) Isospin Amplitudes and Crossing Matrices}\label{app}

In this appendix, we define the isospin conventions, the crossing matrices, and the specific NLO imaginary parts utilized in the main text for $SU(2)$ Chiral Perturbation Theory (ChPT).

The full isospin amplitudes $T^I(s,t,u)$ for pion-pion scattering can be decomposed in terms of a single invariant scattering amplitude $\mathcal{A}(s,t,u)$:
\begin{align}
T^0(s,t,u) &= 3\mathcal{A}(s,t,u) + \mathcal{A}(t,s,u) + \mathcal{A}(u,t,s) \,, \\
T^1(s,t,u) &= \mathcal{A}(t,s,u) - \mathcal{A}(u,t,s) \,, \\
T^2(s,t,u) &= \mathcal{A}(t,s,u) + \mathcal{A}(u,t,s) \,.
\end{align}

 The standard $SU(2)$ isospin crossing matrices, denoted as $C_t$ and $C_u$ in the text, map the $s$-channel isospin vector $T_s = (T^0, T^1, T^2)^T$ to the crossed channels via $T_s^I = \sum_{I'} C_t^{II'} T_t^{I'}$ and $T_s^I = \sum_{I'} C_u^{II'} T_u^{I'}$. They are explicitly given by:
\begin{equation}
C_t = 
\begin{pmatrix}
1/3 & 1 & 5/3 \\
1/3 & 1/2 & -5/6 \\
1/3 & -1/2 & 1/6
\end{pmatrix}
\,, \quad \quad
C_u = 
\begin{pmatrix}
1/3 & -1 & 5/3 \\
-1/3 & 1/2 & 5/6 \\
1/3 & 1/2 & 1/6
\end{pmatrix} \,.
\end{equation}

For the explicit evaluation of the left-hand cut integrals in Section \ref{sec:iso2}, we require the imaginary parts of the physical amplitudes over the right-hand cut ($s \ge 4m_\pi^2$). At NLO in ChPT, these are strictly determined by perturbative unitarity (the optical theorem) from the tree-level amplitudes, yielding:
\begin{align}
\text{Im } T^0(s,t,u) &= \frac{2s^2 - 2sm_\pi^2 + \frac{11}{6}m_\pi^4}{16\pi f_\pi^4} \sqrt{1 - \frac{4m_\pi^2}{s}} \,, \\
\text{Im } T^1(s,t,u) &= \frac{t - u}{96\pi f_\pi^4} (s - 4m_\pi^2) \sqrt{1 - \frac{4m_\pi^2}{s}} \,, \\
\text{Im } T^2(s,t,u) &= \frac{\frac{1}{2}s^2 - 2sm_\pi^2 + \frac{10}{3}m_\pi^4}{16\pi f_\pi^4} \sqrt{1 - \frac{4m_\pi^2}{s}} \,.
\end{align}
where $m_\pi$ is the pion mass and $f_\pi$ is the pion decay constant. These imaginary parts act as the spectral inputs for the algebraic series expansion developed in this work.

\bibliographystyle{JHEP}
\bibliography{references.bib}
\end{document}